\documentclass{article}
\usepackage{graphicx} 

\title{Evolution of long-period compact radio sources driven by winds}
\author{J.E. Horvath, Lucas M. de S\'a, L\'ivia S. Rocha, Gustavo Y. Chinen,\\ Lucas G. Bar\~ao and M\'arcio G. B. de Avellar}
\date{IAG-USP, S\~ao Paulo, Brazil\\foton@iag.usp.br}

\begin{document}

\maketitle
We address in this work the nature and evolution of the long-period compact star sources, which has recently added several unexpected members. The central hypothesis is that particle winds drive their evolution, being an important factor for these relatively old sources. We show the consistency of this picture and remark some unsolved problems and caveats within it.
\bigskip


\vfill\eject
\section{Introduction}
The study of the first detected manifestation of neutron stars (NSs), i.e. radio pulsars, has entered its 6th decade being enlarged and extended very substantially in many aspects. On the one hand, several periodic sources have been discovered in {\it other} bands, reflecting not only the pulsed emission seen in other frequencies, but also other environments and situations harboring neutron stars ``caught in the act''. In addition, a variety of behaviors within the enlarged set of radio sources have been identified, possibly related to neutron stars as well, but rising a number of important unanswered questions (Popov 2023). In short, we are at crossroads in the field, in need of a general classification scheme and seeking concrete models to explain the variety of emission types with a minimum testable set of physical hypotheses.

An excellent example of this type of search is the work of Kaspi (2010) trying to make sense of the different groups of objects. Recent work by Popov (2023) addresses the same problem, adding fast radio bursts (FRBs) to the list. Links between these different groups have been explored (e.g. Yoneyama et al. 2019). However, a major new discovery is that of local long-period radio sources, something quite unexpected and puzzling in many ways (see, for instance, Rea et al. 2024). Although discussion of long-period pulsars/magnetars had already been present in the literature before, such as with their possible association with FRBs (e.g. Wadiasingh et al. 2020 and Beniamini et al. 2020, the discovery of so many sources in such a short period came as a surprise, and the mechanism through which these objects happen to reach such low periods while spinning down at a relatively high rate, as well as their emission mechanism, is still under investigation (see, e.g. Beniamini et al. 2023). We attempt in this work to establish an evolutionary link between the long-period sources and the rest of the population. We shall see below to which extent such a proposal can be supported.

\subsection{Well-established facts and general features about radio pulsars}

Virtually every astrophysicist alive knows the spectacular history of the discovery of pulsars by the Cavendish group at Cambridge, UK, operating the IPS instrument at Mullard (Hewish et al. 1979), mainly due to the persistence and insight of Jocelyn Bell  on the reality of a pulsed persistent source. The prompt publication of the interpretation papers by T. Gold (1969) and F. Pacini (1968) and early work showing the handful of new objects roughly fitting the Gold-Pacini scenario, convinced the community that those pulsars were rotating neutron stars. The pulsed emission stems from a complicated magnetosphere dynamics, which in turn is powered by rotational energy, and only partial solutions have been obtained (Goldreich and Julian 1969).

The basic assumptions of the rotating neutron star picture of pulsars is well-known: provided the free-energy source is the rotation of the object, and assuming a single energy loss mechanism, namely the observed pulsed electromagnetic emission, the dynamics is simply given by

\begin{equation}
    \label{eq:simple_dynamics}
    I {\dot \Omega} = \tau_{em},
\end{equation}

\noindent

in which the electromagnetic torque $\tau_{em} \propto B^{2} \Omega^{3}$ can be obtained by integrating the Poynting vector over a closed surface engulfing the source (Manchester and Taylor 1977). The simplest expression stems from the assumption of a vacuum surrounding the neutron star, but it was promptly realized that the strong induced electric fields would lift charges from the surface and fill the space, a fact that complicated the solution of the magnetosphere. The important points to be remarked here are that a) only the dipole mechanism contributes to the energy loss and spindown; and b) every coefficient in the dynamical eq.(1) is constant. Both assumptions, successful as they are, fail to some extent when the dynamics of actual systems is analyzed.

On the one hand, the very picture of open field lines (Goldrteich and Julian 1969) in the magnetosphere, a widely accepted feature, leads to considering the escape of particles along them, i.e. a wind carrying away energy and angular momentum. This was recognized quite early, and discussed over the year (Michel 1969, Wilson and Rees 1978) (see Kirk et al. 2009 for a general discussion), but eventually did not become part of the ``standard'' description (Manchester and Taylor 1977). However, winds are not only expected on theoretical grounds: in the cases in which a rough estimate of the actual dipole pulse loss was possible, it became clear that it is too small to be responsible for the full rotational energy loss. In addition, a few cases in which a wind is strongly indicated by observations are available, the most notorious being the pulsar PSR B1931+24 (J1933+2421) (Kramer et al. 2006), which shows a large difference in its torque when it switches on and off regularly; but also, as a recent example, PSR J2021+4026, which is the only known variable gamma-ray pulsar (Wang et al. 2018, Rigoselli et al. 2021, Razzano et al. 2023). Magnetar spin-down glitches, in which the object loses angular momentum, have been posited to be caused by episodes of wind mass loss as well (Younes et al. 2023). Finally, quasi-permanent changes in the torque after a glitch (Lyne et al. 2015) in the Crab and other pulsars suggested either a discrete change of the angle between the rotation and magnetic axes (Allen and Horvath 1997), or a permanent decoupling of a component contributing to the moment of inertia (Alpar et al. 1993). These and other indications strongly argue against a constant, unique torque as postulated in the simplest picture of pulsar dynamics.

\section{New kids on the block: other compact systems and the long-period sources}

It has been long established that NSs are born in massive star supernova explosions. A relatively recent consensus is that lower mass stars, in the range of $8-10 \, M_{\odot}$ also form NSs when they undergo electron capture onto an O-Ne-Mg core. The process known as {\it accretion induced collapse} (AIC), either in its single-degenerate or double-degenerate  version, is still uncertain but may contribute too (Tauris et al. 2013, Wang and Liu 2020, Wang et al. 2022). A simple multiplication of the core-collapse supernova rate by the estimated Galactic lifespan gives a large number of NSs $\sim \, 10^{8}$ expected to reside in the Milky Way (Bisnovatyi-Kogan 1992). How many of them are actually observable is a different matter: {\it active} pulsars today could feature $70,000$ objects (Lyne et al. 1985, Dirson et al. 2022), but this number depends on the birth parameters and evolution of magnetic fields and is uncertain by a large factor. It is also thought that at some point the old pulsar population fades away by crossing a ``death line'' (see, e.g., Chen and Ruderman 1993, Ho et al. 2021, Ho et al. 2021)). Keeping this in mind is important, since the new long-period objects may challenge the standard consensus (see below). Nevertheless, the conclusion that most of the Galactic NSs are {\it not} pulsars, but may be ``hidden'' or belong to some of the other groups, is unavoidable. As mentioned before, attempts to unify all NSs have been presented before (Kaspi 2010), but as we shall see, new puzzling detections and unresolved problems are abundant.

Although the impact of the discovery of a physical realization of neutron stars in radio pulsars caused an enormous impact, the zoo of compact star systems related to neutron stars was enlarged in the '70s and beyond, with various classes of sources interpreted as different manifestations of the former within a variety of systems.

The oldest example is the group of {\it X-ray pulsars} (Giacconmi et al. 1971), discovered in the early days of dedicated satellite observations, and is credited to the {\em Uhuru} Team, who observed and characterized the source Cyg X-3 in 1971 (Charles and Seward 1995 and references therein). A more general group of sources was identified and classified as belonging to the Low-Mass X-Ray Binary (LMXB) or High-Mass X-Ray Binary (HMXB) classes depending on the companion mass (solar-type or massive, respectively), and a wealth of information on accretion disks and many other features gathered (Pszota and Kov\'acs 2023). The presence of X-ray pulsars in some binary systems of the LXMB class has been confirmed (Jonker and van der Klis 2001), but most of the latter do not show pulsations (Patruno et al. 2018). The short period {\it millisecond pulsars} (Bisnovatyi-Kogan and Komberg 1974) may in fact be descendants of some type of LMXB.

The second oldest example of these groups we should mention are the {\it Central Compact Objects} (CCOs), quiet NSs in a few supernova remnants with or without low-amplitude pulsations (DeLuca 2017, Mayer and Becker 2021). They are thought to be quite young, as corresponding to their birth in the supernova events, and their emission may be pointed away from us or, alternatively, they may have a small magnetic field/rotation rate. The compact soft X-ray source 1E 161348-5055 makes for a particularly interesting case. It was originally classified solely as an CCO associated to the remnant RCW 103 (De Luca et al. 2006), and from its inception proved to be a puzzling case of a variable source, displaying a periodicity of $6.67\,\mathrm{h}$ and X-ray luminosities ranging from $10^{33}$ to $10^{35}\,\mathrm{erg}\,\mathrm{s}^{-1}$; its unusually high period led to the hypothesis of it being a X-ray binary instead of an isolated rotating NS. In either case however, an extreme magnetic field seemed to be required, and D'Ai et al. (2016) did in fact confirm its nature as a slowly-spinning isolated \textit{magnetar} (discussed below), with field $10^{14}-10^{15}\,\mathrm{G}$. Since then, the mechanism by which it might have reached such a long period has remained uncertain; one early proposal was braking from interaction with fallback material from the remnant (De Luca et al. 2006).

The first member of the group known as {\em Isolated Neutron Stars}, RX J1856.4–3754 is more recent (Walter et al. 1996). The object {\it Geminga} (``it is not there'' in Milanese dialect) remained as a prototype for many years (Caraveo et al. 1996), but it has been now joined by a handful of other examples. The members became known as the {\it Magnificent Seven} or X-ray Dim Isolated Neutron Stars (XDINSs). Further candidate members include {\it Calvera} (Rutledge et al. 2008), and 4XMM J022141.5-735632 (Pires et al. 2022). The XDINSs are relatively close, blackbody-like cooling NSs (Potekhin et al. 2015), and pulsations and non-zero period derivatives were detected over time, proving that they are actually middle-aged objects (Turolla 2009, Ertan et al. 2014). It was expected that their number would be much higher by now, but apparently this is not the case.

A serendipitous detection of {\it Rotating Radio Transients} (RRATs), emitting occasionally isolated pulses of a few ${ms}$ and going silent for days or more was first made in 2006 (McLaughin et al. 2006). Their narrow duty cycles make them virtually invisible most of the time, and therefore their number could be very large. It has been suggested that they may be as numerous as radio pulsars, or even more abundant. When coherent solutions for the sporadic pulses were constructed, they allowed an estimate of some of the period derivatives ${\dot{P}}$, and therefore an estimate of the characteristic spin-down ages. They seem to be not too different from ordinary pulsars in this aspect; although the recent detection of the extremely intermittent radio source PSR J1710-3452, which shows similarities to radio-loud magnetars and is placed close to magnetars in the $P-\dot{P}$ plane, but would feature as a particularly old magnetar, might complicate this picture (Surnis et al. 2023). The suggestion is that, for some reason the ``spark'' leading to a pulse does not operate efficiently, as in the pulsar case (Abhishek et al. 2022). It is possible that this behavior is an extreme example of the {\it nulling} phenomenon seen in many ordinary pulsars. However, we do not know enough about the actual origin of pulses to test these ideas.

Another subgroup of NSs, of particular importance for our discussion, is the {\it magnetar} class, in which
the dominating X-ray emission has been attributed to the existence of a large magnetic field. In fact the X-ray luminosity ${\dot{E}}_{X}$ exceeds the rotational energy reservoir (Esposito et al. 2021) $I\Omega{\dot{\Omega}}$, and makes magnetars objects powered by magnetic energy. It is a matter of debate whether there is a continuum of NSs, in the sense that the magnetic fields form a continuous distribution, or if there is some kind of gap between pulsars and magnetars instead. {\it Transition pulsars}, with magnetic fields as high as some of the identified magnetars, but otherwise showing quite normal radio emission, argue in favor of a continuum in $B$. The (unexpected) existence of ``low-field'' magnetars (Turolla and Esposito 2013) adds strength to the latter hypothesis. It is important to remark that the magnetar group is defined as the NSs for which the magnetic field is the dominant source of energy, relative to rotational energy, regardless of the particular emission mechanism.

The most recent unexpected discovery of long-period radio sources can be seen as a further example of our incomplete knowledge of compact stellar radio sources. A series of detections of progressively longer period pulsar-like (Caleb et al. 2022) and magnetar-like (Hurley-Walker et al. 2022) sources begun a few years ago and is still ongoing. Some of these long-period sources are located below the ``classical'' {\it death line}, and a definite division between them is far from clear, as their position in the $P-{\dot{P}}$ plane does not allow one to establish a neat classification. A couple of examples are located very close to the pulsar-like white dwarf objects AR Scorpii (Buckley et al. 2017) and the recently discovered  J191213.72-441045.1 (Pelisoli et al. 2023). In fact, one of these members (GPM J1839-10) was known since 1988 (Hurley-Walker et al. 2023), but had ``disappeared'' from the research landscape. It is challenging, and the main goal of the present work, to relate these sources to the rest of the population.

\section{Are the long-period sources driven by winds?}

The description of the pulsar wind physics and their evolution is still incomplete, in spite of being the subject of an early research (Michel 1971, Michel 1973, Usov 1975, Coroniti 1990). A useful review by Kirk, Lyubarsky and P\'etri (2009) can be consulted for a general view of the achievements and open problems. While winds may carry ions and pairs in comparable number escaping through the open lines, as expressed above, it has not been easy to address the actual importance of the winds for the object, and its contribution to the energy loss has been neglected usually in favor of the dipole contribution.

The new long-period objects may offer another view of this problem. Since the dipolar loss scales as $\Omega^{3}$ in the dynamical Eq.(1), slow rotators are expected to quench the dipole emission at some point. This is a fact behind all present theories of the ``death line''. However, GPM J1839-10, for instance, is probably below all versions of these ``death lines''. Therefore, the very generation of the pulses is challenged within an orthodox view.

The detailed record of the pulses encourages to think that there are additional factors at work for this group. The time development of the pulses of GPM J1839–10 (Hurley-Walker et al. 2023) shows a very peculiar pattern never seen before, and holds the clue to their exact origin, although they can hardly be labeled as ``canonical''. At this point one may wonder whether additional physics is needed to cope with the pulse generation and track the evolution of the sources (Tong 2023). In terms of evolution, the need for additional explanation for these long-periods had already been recognized at the time of the discovery of 1E 161348-5055; as mentioned before De Luca et al. (2006) proposed an early interaction with fallback material as a braking mechanism, as well as the need for extreme magnetic fields -- these later confirmed for the case of 1E 161348-5055. This mechanism has also been proposed for the more recent GLEAM-X J162759.5-523504.3 (GLEAM J1627 hereon) and PSR J0901-4046  discoveries (Gencali et al. 2023), yielding a possibility that GLEAM J1627 still is surrounded by a disk, and will continue to evolve for the next few $10^5\,\mathrm{yr}$, towards a period of a few $10^3\,\mathrm{s}$ (see also Ronchi et al. 2022) for fallback accretion as a general formation channel for long-period pulsars). Early amplification of the field might also be required to explain the current inferred extreme field of GLEAM J1627 (Suvorov and Melatos 2023).

Motivated by a different type of source (magnetars) showing outbursts, Harding, Contopulos and Kazanas (1999) explored a minimal model of evolution in which the winds were explicitly considered as causing intermittent outflows. The energy loss by the latter was found to scale as the square root of the particle luminosity $L_{P}$ and quadratically with the angular velocity of the object, $\Omega$. A handful of similar works (Allen and Horvath 2000, Alvarez and Carrami\~nana 2004) considered the winds as an additional energy loss mechanism, in some cases even acting continuously as suggested by the earliest pulsar models (Michel 1971, Michel 1973, Usov 1975, Coroniti 1990).However, some winds are still expected to be intermittent, as the sudden and large mass outflows thought to be necessary to sustain this kind of energy loss are hardly continuous; Beniamini et al. (2020), for example, associated the wind mechanism with periodic giant flares of magnetars with very long periods $P\sim 25\,\mathrm{yr}$. The suggestion we want to make here is that the long-period sources may be another example in which winds play an important role. We shall elaborate this idea below.

\subsection{Basics of winds}

A general appraisal of the magnetosphere and winds can be found in Michel (1991). For the simplest model, the dynamical eq. (1) can be written in terms of the period $P$ for a model including the dipole and wind loss mechanisms (Harding,Contopulos and Kazanas 1999, Allen and Horvath 2000), yielding

\begin{equation}
I \dot{P}=4\pi^2\mu^2\frac{(1-\alpha)}{P}+\sqrt{L_P}\mu {\alpha P},
\end{equation}

\noindent where $\mu$ is the magnetic dipole moment of the object, namely $\mu = \frac{B_0r_0^3}{\sqrt{6c^3}}$; and $\alpha$, in an interpretation of eq.(1) as an exact evolution law, is the fraction of the luminosity carried by the wind, such that $L_{tot} = (1-\alpha) {\dot E_{d}} + \alpha{\dot E_{w}}$. Alternatively, $\alpha$ is the wind \textit{duty cycle}, as originally in Harding,Contopulos and Kazanas (1999), defined as the fraction of total ``on'' time; in this sense, eq.(1) describes the \textit{average} energy loss throughout the life of the pulsar. Winds are not expected to be active at all times, and thus $\alpha < 1$.

Within this perspective, the dynamics of a compact rotating source displays a more complex behavior. First of all, it is clear that for each object, there is a transition period separating the epochs of dipole- and wind-dominated evolution (considering $L_{P}$ a fixed quantity), namely

\begin{equation}
    P_\mathrm{tran}= 2\pi\sqrt{\frac{{\mu}(1- \alpha)}{\alpha\sqrt{L_P}}}.
 \label{eq:inflexão}
\end{equation}

As expected, the $P_\mathrm{tran}$ values range from $1$ to $1000\,\mathrm{s}$ depending on the particle luminosity $L_{P}$, a quantity hard to pin down from observations (see below). However, and having some optimism for the future observations, a relationship between the particle luminosity and source quantities can be worked out quite simply by combining the particle wind coefficient of Eq. (2) with Eq. (12) from Allen and Horvath (2000), with the result

\begin{equation}
 L_{P} = {\biggl( {{I\dot{P}(3-n)}\over{{2\alpha \mu P}}}\biggr)}^2.
 \label{eq:Lp}
\end{equation}

In spite of the fact that the braking index $n$ of the long-period objects is not yet known, and constitutes a quite difficult determination for the observed rotation, Eq.(4) converts $L_{P}$ into a potentially observable quantity. As a consistency check for the wind hypothesis, we may look at the known young pulsars with measured braking indices, under the view that they are also affected by some degree of wind braking, a specially attractive possibility for those with low braking indices $n \geq 1$. Table 1 shows some of the results. All the calculated particle luminosities $L_{P}$ for these objects are much higher than the observed X-ray emission, a fact that suggests that particle escape can operate ``silently'', and may still be quite high in old objects. We shall return to this issue below.

\begin{table}[tbhp]
{Table 1. The $L_P$ values calculated with Eq. (4) for a set of pulsars with known braking indices. In all cases $r_0 = 10^6\,\mathrm{cm}$, $I = 10^{45}\,\mathrm{g}\,\mathrm{cm}^2$ and $\alpha = 0.5$ were considered.}

\bigskip
\center
{\begin{tabular}{cccc}

\hline

Pulsar  & $n$  & $\log L_P$ & $\log L_X$\\
{} & {} & $\left(\mathrm{erg}\,\mathrm{s}^{-1}\right)$ & $\left(\mathrm{erg}\,\mathrm{s}^{-1}\right)$ \\

\hline
Crab$^{1}$ & 2.51 & 37.60 & 36.21 \\
PSR$^{1}$ J1846-0258 & 2.16 & 37.84 & 35.23\\
Vela$^{1}$ & 1.4 & 36.70 & 31.48\\
PSR J1734-3333$^{2}$ & 1.1 & 35.27 & 33.53\\

\hline

\end{tabular}}

\small{$^{1}$ Shibata et al. (2016),Hamil (2015), $^{2}$ Shibata et al. (2016) ,Olausenet al. (2010).}

\label{tab:psr1}

\end{table}

Finally, a value of the magnetic field $B$ can be found by inverting the dynamical equation $L_{tot} = (1-\alpha) {\dot E_{d}} + \alpha{\dot E_{w}}$ (see Harding, Contopulos and Kazanas 1999), with the result

\begin{equation}
B=-\frac{\sqrt{6c^3}}{r_0^3}\left(\frac{\sqrt{L_P}\alpha P^2-P\sqrt{L_P\alpha^2P^2+16\pi^2\frac{(1-\alpha)I\dot{P}}{P}}}{8\pi^2(1-\alpha)}\right)  ,
 \label{eq:B}
\end{equation}

\noindent and gives the time spent by the objects in the wind stage, as calculated in this model (Harding,Contopulos and Kazanas 1999, Allen and Horvath 2000), as

\begin{equation}
 \tau_{w} (P) = {\frac{I}{2\alpha \mu \sqrt{L_p}} \ln{\left(1 + \frac{\sqrt{L_p}P^2\alpha}{4\pi^2\mu(1-\alpha)}\right)}}.
 \label{eq:idade}
\end{equation}

\begin{figure}[pb]
\centering
\includegraphics[width=8.2 cm]{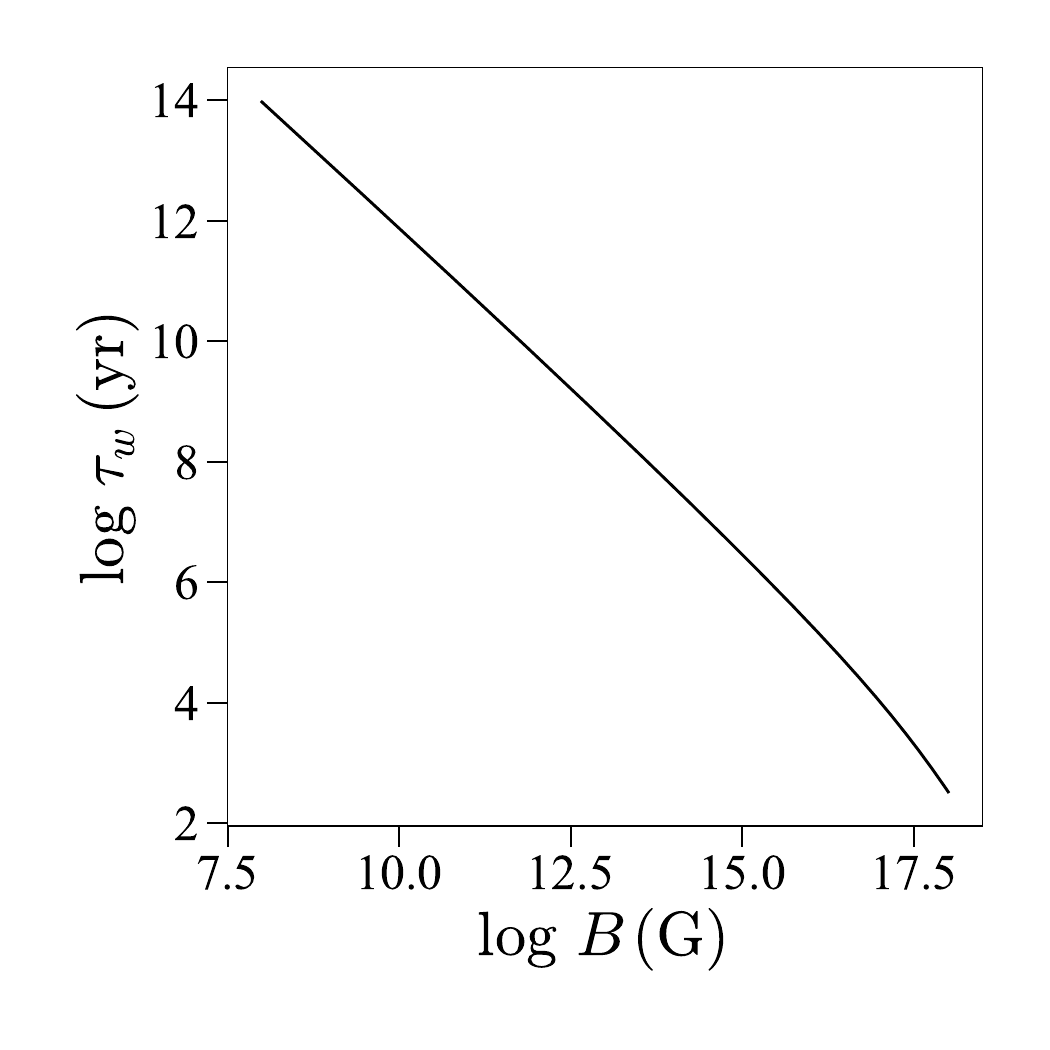}
\caption{Calculated age for a pulsar with $P = 2000 \,\mathrm{s}$, ${\dot P} = 10^{-12}\,\mathrm{s}\,\mathrm{s}^{-1}$ and  $L_{P} = 10^{30}\,\mathrm{erg}\,\mathrm{s}^{-1}$, values close to the ones of the slowest objects in Table \ref{tab:psr2}. The values $\alpha = 0.5$, $r=10^6\,\mathrm{cm}$ e $I=10^{45}\,\mathrm{g}\,\mathrm{cm}^2$ were adopted. note that the objects with $B \sim 10^{12}\,\mathrm{G}$ are possibly as old as the galaxy.\label{Age}}
\end{figure}

In some cases, depending on the value of the magnetic field $B$ and the particle luminosity $L_{P}$, the wind stage largely dominates the total age (i.e., the time after the wind dominance is larger than the initial dipole-driven stage), and $\tau_{w}$ can be considered the age of the object, in particular when its value is in the $\sim\mathrm{Gyr}$ range (Fig. \ref{Age}). The evolution (if any) of the magnetic field, along with the speed at which the neutron star spins, play a crucial role in shaping the electromagnetic surroundings and, in turn, the emissions we observe. By comparing the emissions across a range of wavelengths from different neutron stars, including the long-period sources, we can get a better understanding of the varying magnetospheric conditions found in these intriguing objects, a long-due task in the field.

\subsection{The wind as a source of pulsed emission}

As already mentioned, although the model above describes how a particle wind generates enough torque to potentially explain the spin-down of a pulsar, it makes no claims about how this wind would be observed, if it were to be observable at all. The wind itself does not require the magnetic and rotational axes to be misaligned in order for this torque to exist, but the assumption that the observed pulsed emission has the wind as a source does. Although we will not attempt to directly model the wind emission here, it is worth considering how the pulsar wind changes when we do not assume an axisymmetric system, and whether possible emission mechanisms are minimally compatible with observations.

Emission from the wind zone of pulsars is notoriously difficult to characterize.  As discussed by Kirk, Lyubarsky and P\'etri (2009), little emission is expected in cold, ideal magnetohydrodynamic (MHD), axisymmetric, wind models, wherein the field adopts a split-monopole configuration, particle trajectories are rectilinear and thermal motion is negligible in relation to the expansion suffered by the plasma in the wind region (by a factor of $10^{18}$ in volume), such that neither synchrotron radiation nor bremsstrahlung can be expected. Only inverse Compton scattering of photons, from sources such as the CMB or a binary companion, provide a possible, weak, source of gamma-rays.

More promising for the objects here discussed is the \textit{striped wind} model (Michel 1971, Coroniti 1990). By abandoning the ideal MHD regime, Michel (1971) had already shown that the plasma flow in the equatorial region should evolve into successive bands of cold plasma, each dominated by magnetic field lines originating in alternating poles, separated by a narrow current sheet of hot plasma. Instead of following the equatorial plane, however, this current sheet oscillates around it, with amplitude growing linearly with distance, such that, for a narrow enough belt around the equator, flow resembles expanding spherical current sheets, along the boundaries of which magnetic energy dissipates into particle energy if magnetic reconnection occurs, providing a possible source of synchrotron radiation (Kirk, Skjaeraasen and Gallant 2002).

Due to relativistic beaming, photons originating at any point on the shell will only reach the observer if emitted within a narrow cone of opening $\propto\gamma^{-1}$, $\gamma$ the Lorentz factor of the wind, in relation to the line-of-sight, leading the emission to be point-like and possibly indistinguishable from the pulsar itself at current resolutions. Additionally, because the periodicity of the flow is coupled to the rotation of the pulsar --- in particular, the inversion of the field direction at a fixed radius happens with the rotation period ---, the emission is expected to be pulsed and with a similar profile to the modulation of the flow, if not the same period, as long as certain geometrical constraints are fulfilled.

The resulting profile by Kirk, Skaeraasen and Gallant (2002) shows a pulse and an interpulse, the interval between which depends on the viewing angle in relation to the equatorial plane, where they are symmetrical. Among other idealizations, they highlight that their model assumes emission happening at the exact position of the current sheet, while in reality emission might happen within an extended region, making the pulse more complicated; in this case, their result serves as a Green's function for modeling the entire pulse, which might display structure.

The point-like and pulsed nature of striped wind emission is encouraging for attempting to study the long-period sources under this model. First, the point-like nature allows both dipole and wind emission to be {\it superposed} in the observed signal, as needs to be the case for application of the mixed model. Second, its nature as a periodic signal with possible pulse-substructure leaves room for both the unexpectedly irregular profiles of GPM J1839–10 (Hurley-Walker et al. 2023), PSR J0901-4046 (Caleb et al. 2022) and 1E 161348-5055 2001 data (De Luca et al. 2006) and the more regular profile of GLEAM J1627 (Hurley-Walker et al. 2022), PSR J0250+5854 (Tan et al. 2018) and 1E 161348-5055 2005 data (De Luca et al. 2006) . Lastly, the coupling between pulsar rotation and the wind allows us to continue to employ the measured pulse period as the rotation period.

This last point is of particular importance, because for non-dipole emission mechanisms, the spin and pulse periods need not be the same in general. Given that the wind model still requires the spin period to be known, it is important to check that it is still reasonable to assume the two periods to be equal even in this case; if the radio emission has the wind as a source, than we can say that this is the case for these objects. The similarity between high-energy and radio pulses in the case of the Crab pulsar has been suggested to point towards a common origin in the past (Kirk, Skaeraasen and Gallant 2002), although this remains speculative.

If the radio pulses are not originated in the wind, then there remains a question of what mechanism is responsible for their long duration and irregular profiles. Beyond pulse duration and profile, radio dipole emission has traditionally been thought to rely on pair production, which must compete with photon splitting under intense magnetic fields; this has been proposed as the mechanism behind many high magnetic field pulsars observed in the X-rays being radio-quiet (Baring and Harding 1998). An analogous effect has been shown to emerge from the split-monopole configuration of the field (Hu et al. 2022), which is adopted in the aligned rotator model, but which would otherwise preserve the possibility of striped wind emission from an oblique rotator.

Without a clear picture of the whole emission mechanism, there is no obvious way to connect the spin period and the pulse period. In the scenario suggested here, where the wind features in the spindown of the pulsar, there are many reasons to believe the radio emission is not originated by the dipole mechanism. And while it is hard to associate the striped wind with radio emission, the former has been shown to yield a variety of regular profiles which could possibly reproduce the profile observed in the long-period sources. We thus tentatively retain the assumption that the two periods are the same in order to move forward with the analysis.

\subsection{Status of the long-period sources}

\label{sec:status}

We collected in Tables 2  several relevant quantities needed to address the evolution of long-period sources in the $P-{\dot P}$ plane. Note that we have focused on the upper part of the diagram (roughly defined by ${\dot P} > 10^{-15}\,\mathrm{s}\,\mathrm{s}^{-1}$, because the sources with much lower ${\dot P}$ fall in the region of pulsating white dwarfs (Marsh et al. 2016, Pelisoli et al. 2023) and may be related to them. In addition, unless a hypothesis about a substantial decay of the magnetic field is made, for which there is no firm evidence, the trajectories can not ``bend'' downwards to link the known pulsar-magnetar locus with the former. It is quite realistic to believe that the lower part of the diagram is occupied by pulsating white dwarfs (Coelho and Malheiro 2014, Rea et6 al. 2024), i.e., systems similar to AR Sco and J1912-4410.

\begin{table}[tbhp]
{Table 2. Observed and inferred quantities for the five selected sources located in the upper portion of the $P-{\dot{P}}$ diagram. Upper limits to the derivative were employed to estimate the magnetic fields $B$, although some of their values could be significantly lower.}

\bigskip
\center{\begin{tabular}{cccc}

\hline

\textbf{Object} & $P$ & $\dot{P}$ & $\log B$ \\
{} &  $\left(\mathrm{s}\right)$ &  $\left(\mathrm{s}\,\mathrm{s}^{-1}\right)$ & $\left(\mathrm{G}\right)$ \\

\hline

GPM J1839-10$^{1}$ & 1860		&  $<3.6\times10^{-13}$ & 11.74 \\
GLEAM J1627$^{2}$ & 1090 &  $<6\times10^{-10}$ & 14.84 \\
J0901-4046$^{1}$  & 75.6      &  $2.2\times10^{-13}$  & 14.17 \\
J0250+5854$^{1}$     & 23.54	   &  $2.7\times10^{-14}$  & 12.51 \\
1E 161348-505$^{1}$    & 24012     &  $<1.6\times10^{-9}$ & 15  \\

\hline

\end{tabular}}

\small{$^{1}$ Shibata et al. (2016), Hamil (2015), $^{2}$ Shibata et al. (2016), Olausen et al. (2010).}

\label{tab:psr2}

\end{table}

\begin{figure}[pb]
\centering
\includegraphics[width=8.2 cm]{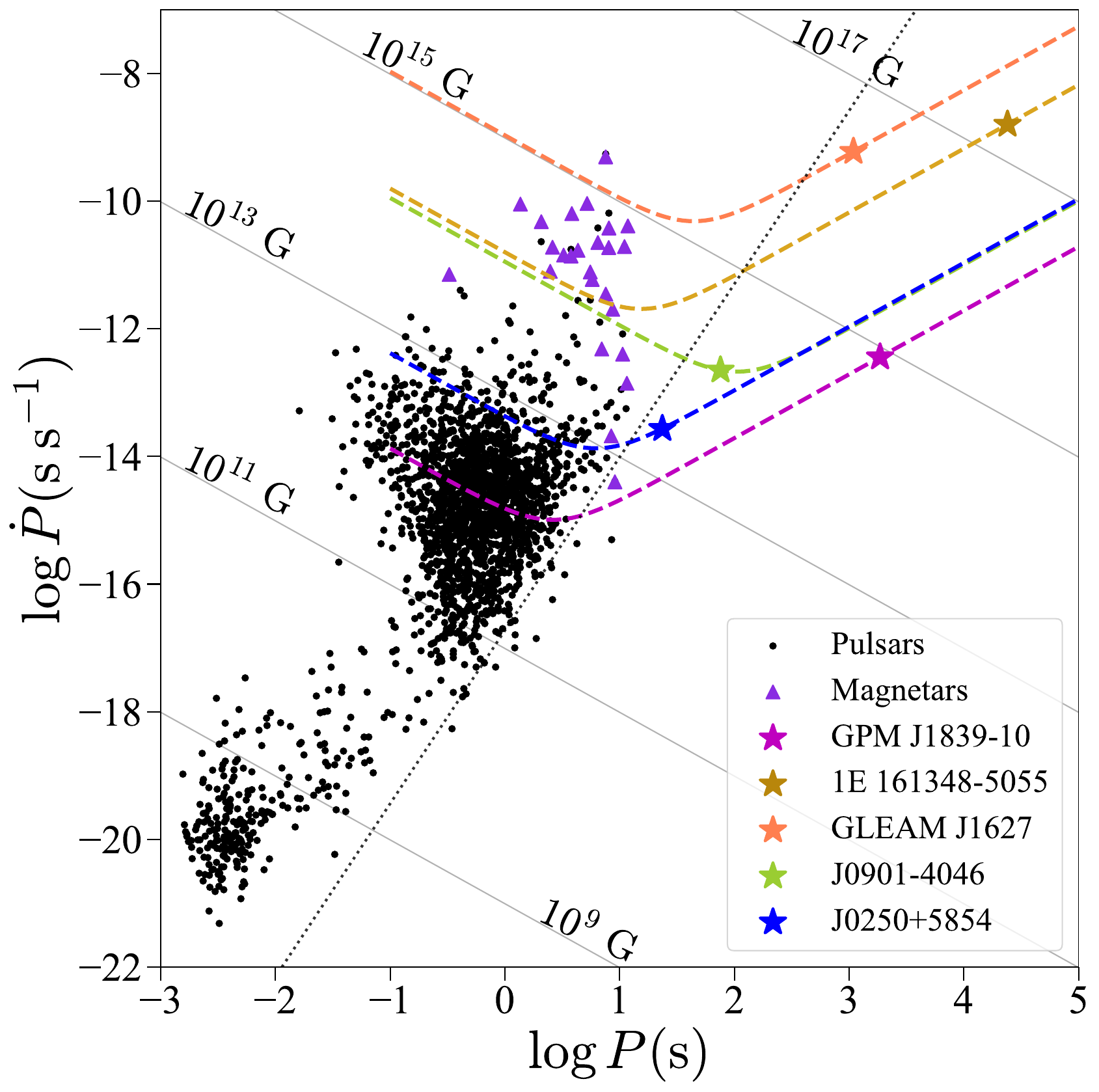}
\caption{The $P-{\dot{P}}$ diagram, showing the position of ordinary pulsars (black circles) and magnetars (purple triangles), as well as the lines of constant magnetic field (solid gray lines) and the death line (black dashed line, from Chen and Ruderman (1993)) in the traditional pure dipole model. The long-period sources are shown as the colored stars, and their possible trajectories in the dipole-wind mixed model shown as the colored dashed lines; the valley separates the dipole- (short periods) and wind- (long periods) dominated epochs. We highlight that the rising portion of the long-period source dipole-wind trajectories do not imply a rising magnetic field as they would in the pure dipole model; their fields are always fixed to the values shown in Tables 2, estimated using eq. \ref{eq:B}. \label{Pulsars}}
\end{figure}

The trajectories of the long-period sources reconstructed from the wind hypothesis are shown in Figure \ref{Pulsars}, for a fiducial $\alpha=0.1$. As stated, there is an intrinsic uncertainty in the particle luminosity $L_{P}$ and the ${\dot P}$ involved in these numbers.

\begin{table}[tbhp]
{Table 2 (Continued) Observed and inferred quantities for the five selected sources located in the upper portion of the $P-{\dot{P}}$ diagram. A value $L_{P} = 10^{34}\,\mathrm{erg}\,\mathrm{s}^{-1}$ was considered as an indicative value for GLEAM J1627. The rotational energy is found simply as $I\Omega^2/2$, and the magnetic energy as $\frac{B^2}{8\pi}V$, $V$ the volume of a spherical shell of external radius $10^6\,\mathrm{cm}$ and thickness $10^5\,\mathrm{cm}$. $L_p^\mathrm{min}$ is the minimum particle luminosity for each object to be in a wind-dominated regime, found as described at the end of Section.}
\bigskip

\center
{\begin{tabular}{ccccc}
\hline

\textbf{Object} & $\log L_{X} $  & $\log L_p^\mathrm{min}$ & $\log E_{rot} $ & $\log E_{mag}$ \\
{} &  $\left(\mathrm{erg}\,\mathrm{s}^{-1}\right)$ & $\left(\mathrm{erg}\,\mathrm{s}^{-1}\right)$ & $\left(\mathrm{erg}\right)$ & $\left(\mathrm{erg}\right)$ \\
\hline

GPM J1839-10$^{1}$ & $\leq$ 33.30  & 26.00 & 40 & 42 \\
GLEAM J1627$^{2}$ & $\leq$ 27.78  & 29.92  & 40 & 46 \\
J0901-4046$^{1}$  & $\leq$ 30.51  & $29.97$ & 38 & 45 \\
J0250+5854$^{1}$  & $\leq$ 33.3   & 30.57 & 42 & 43 \\
1E 161348-505$^{1}$ &  $\leq$ 34   & 26.31 & 43 &  47 \\
\hline

\end{tabular}}

\small{$^{1}$ Shibata et al. (2016), Hamil (2015), $^{2}$ Shibata et al. (2016), Olausen et al. (2010).}

\label{tab:psr3}

\end{table}

This picture leads to identifying the long-period sources with old pulsars/magnetars in which the wind has gained an important role. The most problematic object is GLEAM J1627, which should be the descendant of a very high field object unless its ${\dot P}$ value is far from the upper limit and/or its $L_{P}$ differs substantially from the naive estimation. In fact, this consideration is extensive to the whole group. Note that the comparison between the two last columns of Table 2 allows us to state that the long-period sources are ``magnetars'', if the wind picture holds, just because they are powered by magnetic energy irrespective of the value of the inferred magnetic field.

These results have been obtained using indicative $L_{P}$ values, and this issue deserves clarification. We have seen above that the particle luminosities for the ordinary pulsars in Table 1 are higher than the detected X-ray luminosities $L_{X}$. In addition, the latter is considered to arise mainly from synchrotron emission, not by the effects of the wind. Kirk, Lyubarsky and P\'etri (2009) state that the wind ``glow'' may not be significant if there is not enough matter around and/or the photon field is weak and inverse Compton is inefficient. Therefore, it is not contradictory that the X-ray emission from the long-period sources is undetected (Tables 2), since not only the synchrotron effect is expected to be feeble, but also that enough inverse Compton is not guaranteed. The intensity of the particle luminosity $L_{P}$ can be loosely estimated at best, since a complete and consistent characterization of the winds is not available and many features (i.e. the Lorentz factor of the particles) are poorly known.

Given that the mixed model results in a transition period $L_\mathrm{tran}$ (Eq. 3) between dipole- and wind-dominated epochs for each source in the $P-\dot{P}$ diagram, it is possible to derive one first constraint over the particle luminosity if we impose that the objects be wind-dominated. By fixing $P$ and $\dot{P}$, varying $L_p$ will result in different tracks, for each of which the object will be in a different position relative to the transition between the two regimes. We are thus able to find the $L_p$ for which the object's current period would also be its $P_\mathrm{tran}$ ; this $L_p$ is the minimum particle luminosity for the object to enter the wind regime, reported as $L_p^\mathrm{min}$ in Table 2. Although we do not quantify it, we also point out that the stringent X-ray luminosity limits also simultaneously limit $L_p$ and $\alpha$; the greater the energy carried by the wind, or the more often it is active, the greater would be the expected heating of the surface of the object, which would be reflected in the X-ray luminosity.

\section{Discussion and Conclusions}

We have argued in this work that particle winds, known to be present in ordinary pulsars as a part of the general picture, but rarely considered important, may be the most relevant driving factor to understand the long-period compact star radio sources.

As it stands, it is difficult to state something definitive, but we believe that the first exploration offers a general consistency, and of course a few unexplained features. In addition to theoretical work addressing the specific physics of the pulse generation, observations of the time profile of the latter is a prime element to pin down the features involved. In our separation of the ``upper'' and the ``lower'' sources a deep issue (namely, the decay of the magnetic field) pops up again, because the shape of the trajectories disfavors a strong decay. After many studies, it is clear that quite old sources (Roberts 2012) still show moderately strong fields $B \sim 10^{9}\,\mathrm{G}$, and therefore the decay is, if anything, less extreme than naively expected (Gonz\'alez and Reisengger 2010), at least for isolated sources. A general remark is that the long-period sources are all ``magnetars'', in the sense that they are powered by magnetic energy and not by rotation, independently of the actual value of the magnetic field which  may be comparable to the ones found in ordinary radio pulsars. On the other hand, some observations may be seen as arguing at face value against the suggested wind picture: the central compact object in RCW 103, 1E 161348-5055, has been known for decades, but only recently identified as a magnetar (De Luca et al. 2006). This central source is also radio quiet, but very X-ray bright, which is not easily understood within the presented wind picture.

It is important to remark that other sources have recently contributed to challenging traditional pulsar pictures. The recent observations of novel $\mathrm{TeV}$ emission from the Vela pulsar with H.E.S.S. (2023), for example, join the growing set of new data that challenges the traditional pulsar picture, with the pulsar wind model also being a candidate for explaining the emission in that case. It is also clear that the important issue of pulsating white dwarf systems and their relation with the neutron star group, emphasized by Coelho and Malheiro (2014) and recently revived by Rea et al. (2024), is crucial to construct an overall picture of all sources, as well as the efforts to identify new candidates and the search of counterparts.  Unraveling the evolutionary path of long-period compact radio sources compared to other neutron star groups calls for an integral approach, bringing together observational astronomy, theoretical modeling, and numerical simulations. This effort is not only exciting for shedding light on the mysterious nature of these sources, but also for reshaping our understanding of neutron star astrophysics, pushing what we know into new territories. In summary, these are big news in the field and will keep us busy for a long time.

\section*{Acknowledgments}
This research was funded by FAPESP Agency (S\~ao Paulo State)  grant number 2020/08518-2. L.M.S. received funding from the CNPq (Brazil), grant number 140794/2021-2. L.S.R was funded by Capes Agency (Brazil). J.E.H. acknowledges a long-term Research Scholarship by the CNPq (Brazil) supporting this and other scientific works.

\section{References}
\bigskip\noindent

\bigskip\noindent
Abhishek, Malusare, N., Tanushree, N., Hegde, G., Konar, S. (2022). \textit{Radio pulsar sub-populations (II): The mysterious RRATs. Journal of Astrophysics and Astronomy}, \textbf{43}, 75.

\bigskip\noindent
Allen, M. P., Horvath, J. E. (1997). \textit{Glitches, torque evolution and the dynamics of young pulsars. Monthly Notices of the Royal Astronomical Society}, \textbf{287}, 615-621.

\bigskip\noindent
Allen, M. P., Horvath, J. E. (2000). \textit{Pulsar spin-down with both magnetic dipole and relativistic wind brakings. KITP Conference: Spin, Magnetism and Cooling of Young Neutron Stars}, p. 29.

\bigskip\noindent
Alpar M.A., Pines D. (1993). \textit{Isolated Pulsars Cambridge University Press}, p.17

\bigskip\noindent
\'Alvarez, C., Carrami\~nana, A. (2004). \textit{Monopolar pulsar spin-down. Astronomy \& Astrophysics}, \textbf{414}, pp. 651-658.

\bigskip\noindent
Baring, M. G., Harding, A. K. (1998). \textit{Radio-quiet pulsars with ultrastrong magnetic fields. The Astrophysical Journal}, \textbf{507}, L55.

\bigskip\noindent
Beniamini, P., Wadiasingh, Z., Metzger, B. D. (2020). \textit{Periodicity in recurrent fast radio bursts and the origin of ultralong period magnetars. Monthly Notices of the Royal Astronomical Society}, \textbf{496}, 3390-3401.

\bigskip\noindent
Beniamini, P., Wadiasingh, Z. \textit{et al.} (2023). \textit{Evidence for an abundant old population of Galactic ultra-long period magnetars and implications for fast radio bursts. Monthly Notices of the Royal Astronomical Society}, \textbf{520}, 1872-1894.

\bigskip\noindent
Beniamini, P., Wadiasingh, Z. \textit{et al.} (2023). \textit{Evidence for an abundant old population of Galactic ultra-long period magnetars and implications for fast radio bursts. Monthly Notices of the Royal Astronomical Society}, \textbf{520}, 1872-1894.

\bigskip\noindent
Bisnovatyi-Kogan, G. S. (1992). \textit{The neutron star population in the Galaxy. Symposium-International Astronomical Union}, \textbf{149}, pp. 379-382.

\bigskip\noindent
Bisnovatyi-Kogan, G. S., Komberg, B. V. (1974). \textit{Pulsars and close binary systems. Soviet Astronomy}, \textbf{18}, p. 217.

\bigskip\noindent
Buckley, D. A. H., Meintjes, P. J., Potter, S. B., Marsh, T. R., Gänsicke, B. T. (2017). \textit{Polarimetric evidence of a white dwarf pulsar in the binary system AR Scorpii. Nature Astronomy}, \textbf{1}, 0029.

\bigskip\noindent
Caleb, M., Heywood, I., Rajwade, K. \textit{et al.} (2022). \textit{Discovery of a radio-emitting neutron star with an ultra-long spin period of 76 s. Nature Astronomy}, \textbf{6}, 828-836.

\bigskip\noindent
Caraveo, P. A., Bignami, G. F., Mignani, R., Taff, L. G. (1996). \textit{Geminga in the Space Telescope Era. Astronomy and Astrophysics Supplement}, \textbf{120}, p. 65-68.

\bigskip\noindent
Charles, P. A., Seward, F. D. (1995). \textit{Exploring the X-ray Universe}. CUP Archive.

\bigskip\noindent
Chen, K., Ruderman, M. (1993). \textit{Pulsar death lines and death valley. The Astrophysical Journal}, \textbf{402}, 264-270.

\bigskip\noindent
Coelho, J. G., Malheiro, M. (2014). \textit{Magnetic dipole moment of soft gamma-ray repeaters and anomalous X-ray pulsars described as massive and magnetic white dwarfs. Publications of the Astronomical Society of Japan}, \textbf{66}, 14.

Coroniti, F. V. (1990). \textit{Magnetically striped relativistic magnetohydrodynamic winds-The Crab Nebula revisited. The Astrophysical Journal}, \textbf{349}, pp. 538-545.

\bigskip\noindent
D´A\`i, A. \textit{et al.} (2016). \textit{Evidence for the magnetar nature of 1E 161348 -5055 in RCW 103. Monthly Notices of the Royal Astronomical Society}, \textbf{463}, 2394-2404.

\bigskip\noindent
De Luca, A., Caraveo, P. A., Mereghetti, S., Tiengo, A., Bignami, G. F. (2006). \textit{A long-period, violently variable X-ray source in a young supernova remnant. Science}, \textbf{313}, 814-817.

\bigskip\noindent
De Luca, A. (2017). \textit{Central compact objects in supernova remnants. Journal of Physics: Conference Series}, \textbf{932}, 012006.

\bigskip\noindent
Dirson, L., P\'etri, J., Mitra, D. (2022). \textit{The Galactic population of canonical pulsars. Astronomy \& Astrophysics}, \textbf{667}, A82.

\bigskip\noindent
Ertan, U., Caliskan, S., Benli, O., Alpar, M. A. (2014). \textit{Long-term evolution of dim isolated neutron stars. MNRAS}, \textbf{444}, pp. 1559-1565.

\bigskip\noindent
Esposito, P., Rea, N., Israel, G. L. (2021). \textit{Magnetars: a short review and some sparse considerations. Timing Neutron Stars: Pulsations, Oscillations and Explosions}, 97-142.

\bigskip\noindent
Gencali, A. A., Ertan, U., Alpar, M. A. (2022). \textit{Evolution of the long-period pulsar GLEAM-X J162759. 5–523504.3. MNRAS Letters}, \textbf{513}, L68-L71.

\bigskip\noindent
Gencali, A. A., Ertan,U., Alpar, M. A. (2023). \textit{Evolution of the long-period pulsar PSR J0901 4046. MNRAS Letters}, \textbf{520}, L11-L15.

\bigskip\noindent
Giacconi, R., Gursky, H., Kellogg, E., Schreier, E., Tananbaum, H. (1971). \textit{Discovery of periodic X-ray pulsations in Centaurus X-3 from Uhuru. Astrophysical Journal}, \textbf{167}, p. L67.

\bigskip\noindent
Gold, T. (1969). \textit{Rotating neutron stars and the nature of pulsars. Nature}, \textbf{221}, 25-27.

\bigskip\noindent
Goldreich, P., Julian, W. H. (1969). \textit{Pulsar electrodynamics. Astrophysical Journal}, \textbf{157}, p. 869.

\bigskip\noindent
Gonzalez, D., Reisenegger, A. (2010). \textit{Internal heating of old neutron stars: contrasting different mechanisms. Astronomy \& Astrophysics}, \textbf{522}, A16.

\bigskip\noindent
Jonker, P. G., van der Klis, M. (2001). \textit{Discovery of an X-ray pulsar in the low-mass X-ray binary 2A 1822–371. The Astrophysical Journal}, \textbf{553}, L43.

\bigskip\noindent
Hamil, O., \textit{et al.} (2015). \textit{Braking index of isolated pulsars. Physical Review D}, \textbf{91}, 063007.

\bigskip\noindent
Harding, A. K., Contopoulos, I., Kazanas, D. (1999). \textit{Magnetar spin-down. The Astrophysical Journal}, \textbf{525}, L125.

\bigskip\noindent
Hewish, A., Bell, S. J. \textit{et al.} (1979). \textit{Observation of a rapidly pulsating radio source. In A Source Book in Astronomy and Astrophysics}, \textbf{1900–1975}, pp. 498-504. Harvard University Press.

\bigskip\noindent
Ho, W. C. (2012). \textit{Central compact objects and their magnetic fields. Proceedings of the International Astronomical Union}, \textbf{8}, 101-106.

\bigskip\noindent
Ho, W. C., Zhao, Y., Heinke, C. O. \textit{et al.} (2021). \textit{X-ray bounds on cooling, composition, and magnetic field of the Cassiopeia A neutron star and young central compact objects. Monthly Notices of the Royal Astronomical Society}, \textbf{506}, 5015-5029.

\bigskip\noindent
Hu, K., Baring, M. G., Harding, A. K., Wadiasingh, Z. (2022). \textit{High-energy photon opacity in the twisted magnetospheres of magnetars. The Astrophysical Journal}, \textbf{940}, 91.

\bigskip\noindent
Hurley-Walker, N., Zhang, X., Bahramian, A. \textit{et al.} (2022). \textit{A radio transient with unusually slow periodic emission. Nature}, \textbf{601}, pp. 526-530.

\bigskip\noindent
Hurley-Walker, N., Rea, \textit{et sl.} (2023). \textit{A long-period radio transient active for three decades. Nature}, \textbf{619}, pp. 487-490.

\bigskip\noindent
Kaspi, Victoria M. (2010). \textit{"Grand unification of neutron stars." Proceedings of the National Academy of Sciences} \textbf{107.16}  7147-7152.

\bigskip\noindent
Kirk, J. G., Skjæraasen, O., Gallant, Y. A. (2002). \textit{Pulsed radiation from neutron star winds. Astronomy \& Astrophysics}, \textbf{388}, L29-L32.

\bigskip\noindent
Kirk, J. G., Lyubarsky, Y., Petri, J. (2009). \textit{The theory of pulsar winds and nebulae. Neutron stars and pulsars}, 421-450.

\bigskip\noindent
Kramer, M., Lyne, A. G., O'Brien, J. T., Jordan, C. A., Lorimer, D. R. (2006). \textit{A periodically active pulsar giving insight into magnetospheric physics. Science}, \textbf{312}, 549-551.

\bigskip\noindent
Lyne, A. G., Manchester, R. N., Taylor, J. H. (1985). \textit{The galactic population of pulsars. Monthly Notices of the Royal Astronomical Society}, \textbf{213}, 613-639.

\bigskip\noindent
Lyne, A. G., Jordan, C. A., Graham-Smith, F. \textit{et al.} (2015). \textit{45 years of rotation of the Crab pulsar. Monthly Notices of the Royal Astronomical Society}, \textbf{446}, 857-864.

\bigskip\noindent
Manchester, R. N., Taylor, J. H. (1977). \textit{Pulsars.}

\bigskip\noindent
Marsh, T. R., Gänsicke, B. T. \textit{et al.} (2016). \textit{A radio-pulsing white dwarf binary star. Nature}, \textbf{537}, pp. 374-377.

\bigskip\noindent
Mayer, M. G., Becker, W. (2021). \textit{A kinematic study of central compact objects and their host supernova remnants. Astronomy \& Astrophysics}, \textbf{651}, A40.

\bigskip\noindent
McLaughlin, M. A., Lyne, A. G., Lorimer, D. R. \textit{et al.} (2006). \textit{Transient radio bursts from rotating neutron stars. Nature}, \textbf{439}, pp. 817-820..

\bigskip\noindent
Michel, F. C. (1969). \textit{Pulsar Stellar Wind Torques: NP 0532. In Bulletin of the American Astronomical Society}, \textbf{1}, p. 354.

\bigskip\noindent
Michel, F. C. (1971). \textit{Coherent neutral sheet radiation from pulsars. Comments on Astrophysics and Space Physics}, \textbf{3}, p. 80.

\bigskip\noindent
Michel, F. C. (1973). \textit{Rotating magnetospheres: an exact 3-D solution. The Astrophysical Journal}, \textbf{180}, L133.

\bigskip\noindent
Michel, F. C. (1991). \textit{Theory of neutron star magnetospheres. University of Chicago Press}.

\bigskip\noindent
Olausen, S. A., Kaspi, V. M., Lyne, A. G., Kramer, M. (2010). \textit{Xmm-newton x-ray observation of the high-magnetic-field radio pulsar psr j1734–3333. The Astrophysical Journal}, \textbf{725}, 985.

\bigskip\noindent
Pacini, F. (1968). \textit{Rotating neutron stars, pulsars and supernova remnants. Nature}, \textbf{219}, 145-146.

\bigskip\noindent
Patruno, A., Wette, K., Messenger, C. (2018). \textit{A Deep Pulse Search in 11 Low Mass X-Ray Binaries. The Astrophysical Journal}, \textbf{859}, 112.

\bigskip\noindent
Pelisoli, I., Marsh, T. R., Buckley, D. A. \textit{et al.} (2023). \textit{A 5.3-min-period pulsing white dwarf in a binary detected from radio to X-rays. Nature Astronomy}, pp. 1-12.

\bigskip\noindent
Pires, A. M. \textit{et al.} (2022). \textit{XMM-Newton and SRG/eROSITA observations of the isolated neutron star candidate 4XMM J022141.5-735632. Astronomy \& Astrophysics}, \textbf{666}, A148.

\bigskip\noindent
Popov, S. B. (2023). \textit{The zoo of isolated neutron stars. Universe} \textbf{9}, 273.

\bigskip\noindent
Potekhin, A. Y., Pons, J. A., Page, D. (2015). \textit{Neutron stars—cooling and transport. Space Science Reviews}, \textbf{191}, 239-291.

\bigskip\noindent
Pszota, G., Kovacs, E. (2023). \textit{X-ray Spectroscopic Study of Low-Mass X-ray Binaries: A Review of Recent Progress via the Example of GX 339-4. Universe}, \textbf{9}, 404.

\bigskip\noindent
Razzano, M., Fiori, A., Parkinson, P. S. \textit{et al.} (2023). \textit{Multiwavelength observations of PSR J2021+ 4026 across a mode change reveal a phase shift in its X-ray emission. Astronomy \& Astrophysics}, \textbf{676}, A91.

\bigskip\noindent
Rea, N., Hurley-Walker \textit{et al.} (2024). \textit{long-period Radio Pulsars: Population Study in the Neutron Star and White Dwarf Rotating Dipole Scenarios. The Astrophysical Journal}, \textbf{961}, 214.

\bigskip\noindent
Rigoselli, M., Mereghetti, S., Taverna, R., Turolla, R., De Grandis, D. (2021). \textit{Strongly pulsed thermal X-rays from a single extended hot spot on PSR J2021+ 4026. Astronomy \& Astrophysics}, \textbf{646}, A117.

\bigskip\noindent
Roberts, M. S. (2012). \textit{Surrounded by spiders! New black widows and redbacks in the Galactic field. Proceedings of the International Astronomical Union}, \textbf{8}, 127-132.

\bigskip\noindent
Ronchi, M., Rea, N., Graber, V., Hurley-Walker, N. (2022). \textit{Long-period pulsars as possible outcomes of supernova fallback accretion. The Astrophysical Journal}, \textbf{934}, 184.

\bigskip\noindent
Rutledge, R. E., Fox, D. B., Shevchuk, A. H. (2008). \textit{Discovery of an Isolated compact object at high galactic latitude. The Astrophysical Journal}, \textbf{672}, 1137.

\bigskip\noindent
Shibata, S., Watanabe, E., Yatsu, Y., Enoto, T., Bamba, A. (2016). \textit{X-ray and rotational luminosity correlation and magnetic heating of radio pulsars. The Astrophysical Journal}, \textbf{833}, 59.

\bigskip\noindent
Surnis, M. P., Rajwade, K. M., Stappers, B. W. \textit{et al.} (2023). \textit{Discovery of an extremely intermittent periodic radio source. Monthly Notices of the Royal Astronomical Society: Letters}, \textbf{526}, L143-L148.

\bigskip\noindent
Suvorov, A. G., Melatos, A. (2023). \textit{Evolutionary implications of a magnetar interpretation for GLEAM-X J162759. 5–523504.3. Monthly Notices of the Royal Astronomical Society}, \textbf{520}, pp. 1590-1600.

\bigskip\noindent
Tan, C. M., Bassa, C. G. \textit{et al.} (2018). \textit{LOFAR Discovery of a 23.5 s Radio Pulsar. The Astrophysical Journal}, \textbf{866}, p. 54.

\bigskip\noindent
Tauris, T. M., Sanyal, D., Yoon, S. C., Langer, N. (2013). \textit{Evolution towards and beyond accretion-induced collapse of massive white dwarfs and formation of millisecond pulsars. Astronomy \& Astrophysics}, \textbf{558}, A39.

\bigskip\noindent
Tong, H. (2023). \textit{On the Nature of Long Period Radio Pulsar GPM J1839-10: Death Line and Pulse Width. Research in Astronomy and Astrophysics}, \textbf{23}, 125018.

\bigskip\noindent
Turolla, R. (2009). \textit{Isolated neutron stars: the challenge of simplicity. Neutron stars and pulsars, Springer Berlin Heidelberg.}, pp. 141-163.

\bigskip\noindent
Turolla, R., Esposito, P. (2013). \textit{Low-magnetic-field magnetars. International Journal of Modern Physics D}, \textbf{22}, 1330024.

\bigskip\noindent
Usov, V. V. (1975). \textit{Wave zone structure of NP 0532 and infrared radiation excess of Crab Nebula. Astrophysics and Space Science}, \textbf{32}, pp. 375-377.

\bigskip\noindent
Collaboration, T. H. \textit{et al.} (2023). \textit{Discovery of a Radiation Component from the Vela Pulsar Reaching 20 Teraelectronvolts. arXiv preprint arXiv:2310.06181.}

\bigskip\noindent
Wadiasingh, Z., Beniamini \textit{et al.} (2020). \textit{The fast radio burst luminosity function and death line in the low-twist magnetar model. The Astrophysical Journal}, \textbf{891}, 82.

\bigskip\noindent
Walter, F. M., Wolk, S. J., Neuhäuser, R. (1996). \textit{Discovery of a nearby isolated neutron star. Nature}, \textbf{379}, 233-235.

\bigskip\noindent
Wang, H. H., Takata, J., Hu, C. P., Lin, L. C. C., Zhao, J. (2018). \textit{X-ray study of Variable Gamma-ray Pulsar PSR J2021+ 4026. The Astrophysical Journal}, \textbf{56}, 98.

\bigskip\noindent
Wang, B., Liu, D. (2020). \textit{The formation of neutron star systems through accretion-induced collapse in white-dwarf binaries. Research in Astronomy and Astrophysics}, \textbf{20}, 135.

\bigskip\noindent
Wang, B., Liu, D., Chen, H. (2022). \textit{Formation of millisecond pulsars with long orbital periods by accretion-induced collapse of white dwarfs. Monthly Notices of the Royal Astronomical Society}, \textbf{510}, 6011-6021.

\bigskip\noindent
Wilson, D. B., Rees, M. J. (1978). \textit{Induced Compton scattering in pulsar winds. Monthly Notices of the Royal Astronomical Society}, \textbf{185}, 297-304.

\bigskip\noindent
Yoneyama, T., Hayashida, K., Nakajima, H., Matsumoto, H. (2019). \textit{Unification of strongly magnetized neutron stars with regard to X‐ray emission from hot spots. Astronomische Nachrichten} \textbf{340}, 221-225.

\bigskip\noindent
Younes, G., Baring, M. G., Harding, A. K. \textit{et al.} (2023). \textit{Magnetar spin-down glitch clearing the way for FRB-like bursts and a pulsed radio episode. Nature Astronomy}, \textbf{7}, 339-350.

\end{document}